\documentclass{article}

\usepackage[nonatbib, preprint]{neurips_2023}

\usepackage[utf8]{inputenc} 
\usepackage[T1]{fontenc}    
\usepackage{hyperref}       
\usepackage{url}            
\usepackage{booktabs}       
\usepackage{amsfonts}       
\usepackage{nicefrac}       
\usepackage{microtype}      
\usepackage{xcolor}         
\usepackage[pdftex]{graphicx}
\usepackage{amsmath}

\title{DiffDenoise: Self-Supervised Medical Image Denoising with Conditional Diffusion Models}

\author{%
Başar Demir$^1$\thanks{Equal contribution.}\hspace{0.5em}\thanks{The work conducted by the first author was carried out during an internship at United Imaging Intelligence.} \quad Yikang Liu$^2$\footnotemark[1] \quad Xiao Chen$^2$ \quad Eric Z. Chen$^2$ \quad Lin Zhao$^2$ \\
\textbf{Boris Mailhe}$^2$ \quad \textbf{Terrence Chen}$^2$ \quad \textbf{Shanhui Sun}$^2$ \\
$^1$University of North Carolina at Chapel Hill, Chapel Hill, NC \\
\quad $^2$United Imaging Intelligence, Boston, MA \\
\texttt{basardemir1@gmail.com} \\
\texttt{\{yikang.liu, xiao.chen01, zhang.chen, lin.zhao01,}\\
\texttt{boris.mailhe, terrence.chen, shanhui.sun\}@uii-ai.com}
}

\begin{document}

\maketitle

\begin{abstract}
Many self-supervised denoising approaches have been proposed in recent years. However, these methods tend to overly smooth images, resulting in the loss of fine structures that are essential for medical applications. In this paper, we propose DiffDenoise, a powerful self-supervised denoising approach tailored for medical images, designed to preserve high-frequency details. Our approach comprises three stages. First, we train a diffusion model on noisy images, using the outputs of a pretrained Blind-Spot Network as conditioning inputs. Next, we introduce a novel stabilized reverse sampling technique, which generates clean images by averaging diffusion sampling outputs initialized with a pair of symmetric noises. Finally, we train a supervised denoising network using noisy images paired with the denoised outputs generated by the diffusion model. Our results demonstrate that DiffDenoise outperforms existing state-of-the-art methods in both synthetic and real-world medical image denoising tasks. We provide both a theoretical foundation and practical insights, demonstrating the method’s effectiveness across various medical imaging modalities and anatomical structures.
\end{abstract}

\section{Introduction}

\label{sec:intro}
Medical imaging is invaluable for diagnosing diseases and monitoring their progression without invasive procedures. However, the complex nature of the human body, along with limitations in imaging technology, can introduce artifacts and noise into the images. Improving image quality often requires advanced hardware, longer acquisition times, or higher radiation doses, all of which come with significant costs and potential health implications for patients. Given these challenges, developing effective denoising methods as a post-processing step is highly desirable.

Image denoising, an inverse problem without a unique solution, has been extensively studied in the medical and natural image domains. Conventional methods, including non-local means \cite{nonlocal-means} and total variation minimization \cite{totalvariation}, as well as wavelet-based techniques \cite{wavelet1, wavelet2} and self-similarity metric-based approaches \cite{selfsim}, have been proposed to address this challenge. However, these methods often rely on predefined assumptions and lack generalization to various noise assumptions, resulting in poor performance. 

Recent advancements in deep learning have enabled powerful denoising methods that learn to remove noise from pairs of noisy and clean images \cite{zhang2017beyond}. Supervised methods, such as NAFNet \cite{nafnet}, Restormer \cite{zamir2022restormer}, and Uformer \cite{wang2022uformer}, achieve state-of-the-art performance, offering both fast inference and high restoration capability. In the medical domain, acquiring clean ground-truth images is often impractical or impossible due to high costs and radiation exposure risks for patients, which can restrict the scope of supervised methods.

To address the limitations of supervised learning, self-supervised denoising techniques that enable training without clean reference images have been explored. These methods, such as Noise2Noise \cite{noise2noise}, Noise2Void \cite{krull2019noise2void}, and Blind-Spot Network (BSN) based methods \cite{wang2022blind2unblind, Lee_2022_CVPR, Jang_2023_ICCV}, rely on the statistical properties of noise within the images themselves, allowing models to learn effective denoising from only noisy inputs. Self-supervised methods have shown promise in medical imaging settings, as they leverage existing noisy data and bypass the need for clean ground-truth images, making them feasible for medical domains \cite{xu2021deformed2self, fadnavis2020patch2self}. Despite this progress, challenges remain, as current self-supervised approaches often produce overly smooth results and struggle to preserve high-frequency details crucial for accurate diagnosis in medical imaging, due to ineffective use of information in blind spots.

Another line of research has focused on diffusion denoising methods, which have shown great potential for image generation \cite{ho2020denoising, song2020denoising}. Diffusion models are powerful tools for modeling data distributions, learning to reverse the diffusion process by iteratively transforming noise into detailed, high-quality images, making them ideal for realistic image synthesis and precise restoration tasks \cite{xia2023diffir, fei2023generative, zhu2023denoising}. However, self-supervised image denoising with diffusion models is under-explored. 

\noindent In this paper, we propose \textbf{DiffDenoise}, a self-supervised denoising method tailored for medical images using diffusion models. \textbf{Our primary contributions are:}
\begin{itemize}
    \item We demonstrate, for the first time, that sampling a diffusion model trained on noisy images with strong, noiseless conditions (e.g., BSN outputs) effectively recovers fine structural details without reintroducing noise.
    \item We propose a novel technique, stabilized reverse diffusion sampling, to facilitate robust convergence to clean images during the diffusion process.
    \item We further show that supervised training of a network on noisy images and their corresponding denoised outputs from the diffusion model significantly improves denoising performance and accelerates inference speed.
    \item We conduct extensive experiments on both synthetic and real-world medical image denoising datasets, demonstrating that our model outperforms recent state-of-the-art self-supervised denoising methods, delivering sharper results with greater detail.
\end{itemize}

\begin{figure}
    \centering
    \includegraphics[width=0.5\linewidth]{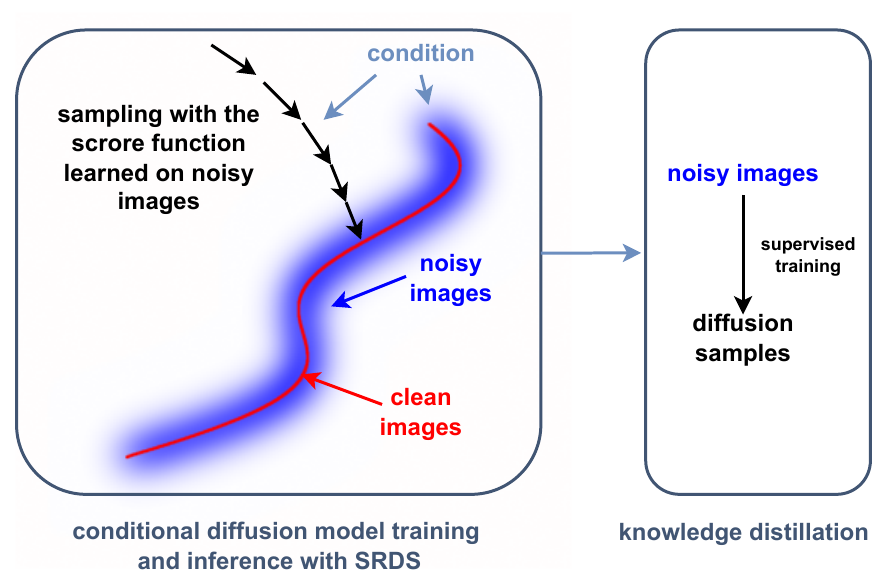}
    \caption{DiffDenoise, a self-supervised denoising method, involves training a diffusion model on noisy images with strong conditioning, sampling clean images with stabilized reverse diffusion sampling, and knowledge distillation to suppress hallucination and accelerate inference.}
    \label{fig:teaser}
\end{figure}

\section{Related Work}
\label{sec:relatedwork}

\subsection{Self-Supervised Denoising}
 Some self-supervised denoising methods assume specific noise distributions, such as Gaussian (\cite{metzler2018unsupervised, zhang2022idr}) or more general exponential family distributions (\cite{kim2021noise2score, kim2022noise}), while some rely on the prior knowledge of noise distribution (\cite{laine2019high,pang2021recorrupted,kim2022noise,xu2020noisy}). These approaches are less applicable for denoising medical images where noise distributions can be intricate and hard to estimate in the image domain. 
 
 Another line of research (\cite{krull2019noise2void, quan2020self2self, wang2022blind2unblind, lee2022ap, zhang2023mm,jang2024puca}) assumes that noise is statistically independent across different dimensions (i.e. pixels), whereas the true signal is correlated (\cite{batson2019noise2self}). These methods thus regress a pixel's noisy value from its neighboring pixels, which have independent noise, thereby achieving denoising without a clean reference. This is typically done either by masking the target pixels in the input image (\cite{krull2019noise2void, quan2020self2self, wang2022blind2unblind, huang2021neighbor2neighbor}) or by using a specialized network architecture known as a "blind-spot network" (\cite{batson2019noise2self, lee2022ap, zhang2023mm}), which is based on the \textit{J}-invariance principle, ensuring that the prediction of a pixel relies exclusively on its surrounding pixels while excluding itself. Although these methods eliminate the need for statistical noise models and have achieved state-of-the-art (SOTA) performance across various datasets (\cite{jang2024puca, li2024tbsn}), they face a notable drawback: excluding pixels within blind spots can lead to the loss of fine details in the resulting denoised images. Some prior works have attempted to address this issue with a regularization loss that incorporates denoising results over full images (\cite{huang2021neighbor2neighbor, wang2022blind2unblind}), but they remain unable to directly recover information from the full images (\cite{zou2023iterative}). Additionally, noise in real-world images often exhibits spatial correlations. To tackle this, recent methods (\cite{lee2022ap, wang2023lg, jang2024puca}) use more distant neighbors to predict the value of the pixel being denoised (i.e., using larger blind spots), under the assumption that noise in these distant pixels is less correlated. However, clean signals in distant pixels are also less correlated, which can result in a greater loss of image information in the denoised results. To mitigate this issue, previous studies have adopted strategies such as using networks with larger receptive fields (\cite{wang2023lg, jang2024puca}), integrating predictions from multiple masked convolution kernels (\cite{zhang2023mm}), or adjusting the blind spot size based on local texture characteristics (\cite{li2023spatially}).

\subsection{Generative Diffusion Models}
Generative diffusion models have become fundamental in unsupervised and self-supervised learning, demonstrating success in various image synthesis and enhancement tasks. These models function within a probabilistic framework, learning to reverse a defined noise-adding process, typically structured as a Markov chain of iterative denoising steps \cite{sohl2015deep, ho2020denoising}. To improve sampling efficiency and output quality, recent research has introduced alternative sampling strategies, including non-Markovian approaches \cite{song2020denoising}.

Conditional diffusion models have significantly expanded the capabilities of diffusion-based techniques, particularly in the field of guided sampling. A notable advancement was introduced by \cite{dhariwal2021diffusion}, which introduced classifier-based guidance to direct the sampling process. This concept was further refined in \cite{ho2022classifier}, enabling models to directly approximate the score function of a conditional distribution from samples of the joint distribution. This development has greatly enhanced the flexibility and effectiveness of these models across a wide range of conditional generation tasks.

In certain scenarios, such as ours, obtaining high-quality datasets is not feasible. The Ambient Diffusion Model \cite{daras2024ambient} addresses this limitation by enabling the use of corrupted training datasets to generate samples from the distribution of the uncorrupted dataset. A subsequent study expanded on this approach to solve inverse problems, including MRI reconstruction \cite{aali2024ambient}. However, their method is restricted to linear corruptions and operates exclusively in frequency domain.

The potential of diffusion models in medical image denoising remains underexplored. The pioneering work in this area \cite{xiang2023ddm} proposed a denoising pipeline limited to specific MRI acquisitions (‘diffusion MRI'). Additionally, \cite{pfaff2024no} conducts an analysis of denoising quality and image fidelity in diffusion based medical denoising strategies.

\section{Methodology}
\label{sec:methodology}

\begin{figure*}
    \centering
    \includegraphics[width=\linewidth]{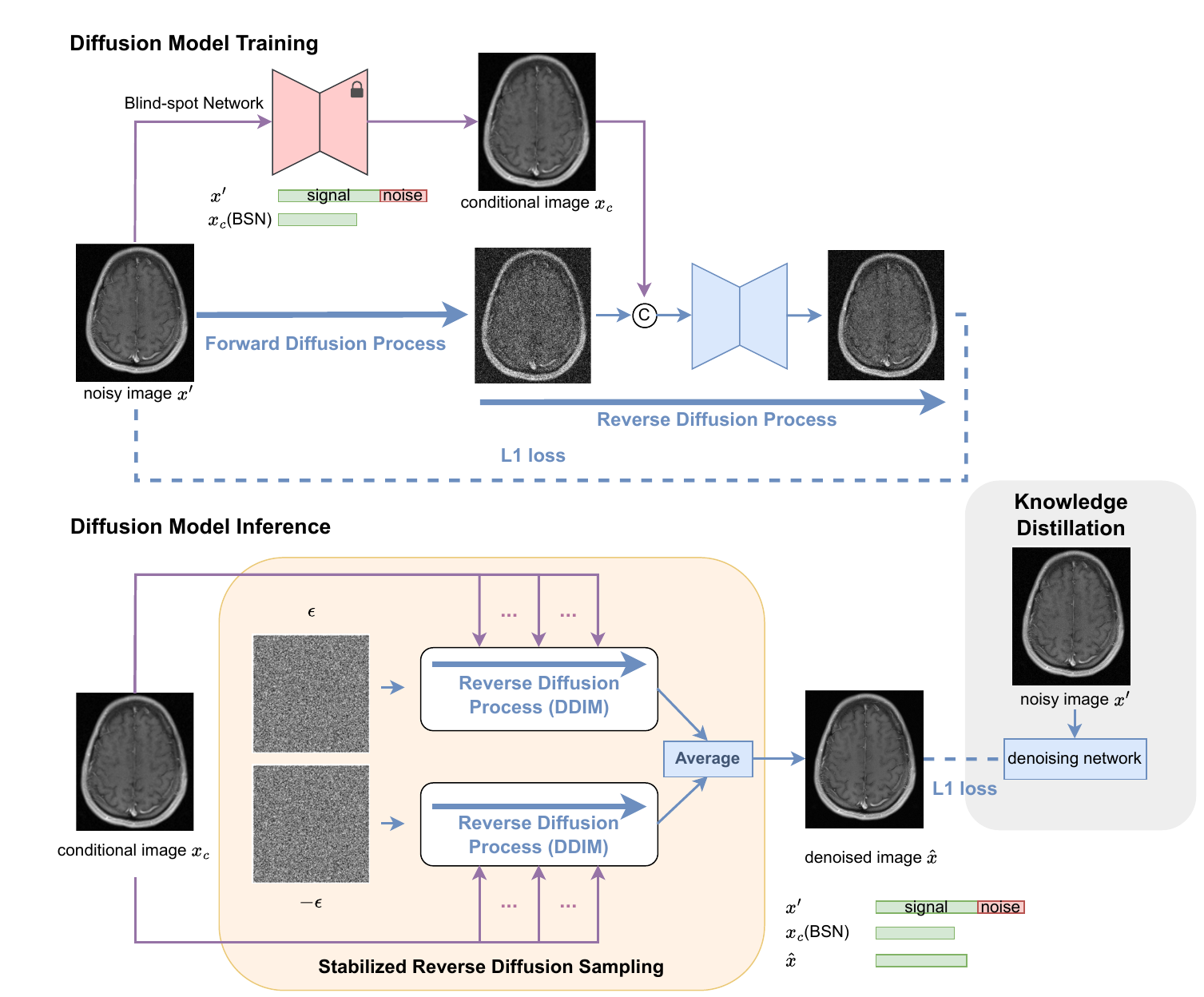}
    \caption{\textbf{Overview of our method.} Our self-supervised denoising method consists of three phases. First, we train a  diffusion model on noisy images, using the outputs of a pretrained Blind-Spot Network as the conditioning input. Then, during inference, we apply our novel stabilized reverse sampling technique that generates clean images by averaging diffusion sampling outputs initialized with a pair of symmetric noises. Finally, we train a supervised denoising network using noisy images paired with the denoised outputs generated by the diffusion model.}
    \label{fig:enter-label}
\end{figure*}

\subsection{Overview}
Our aim is to recover the clean medical images \( x \) from noisy observations \( x' = x + n \), where the noise \( n \) follows an unknown distribution and is not necessarily independent and identically distributed (i.i.d.) across pixels. Section \ref{sec:theo} presents the theoretical foundation of our proposed method, which utilizes a diffusion model to learn the distribution of \( x \) on a dataset of \( x' \), given a strong conditioning variable \( x_c \) that retains most of the information in \( x \) while containing less noise than \( x' \). Section \ref{sec:cond-diffusion} introduces the conditional diffusion models and Section \ref{sec:condition} explains our conditional design. Section \ref{sec:SRDS} proposes a method to stabilize the reverse diffusion sampling of \( p(x | x_c) \). Finally, since diffusion-denoised images may have inconsistencies (due to the nature of the generative process and potential hallucinations) with the input noisy image, Section \ref{sec:knowledge-distillation} introduces a separate neural network to map the noisy image to the diffusion-denoised images, with the network's output serving as the final denoised result of our method.


\subsection{Theoretical Framework}\label{sec:theo}
In medical imaging, for a given signal intensity, many types of noise typically exhibit distributions that peak at zero, are symmetric around zero, and decrease monotonically as they diverge in any direction (e.g. Gaussian or Poisson distribution) \cite{gravel2004method}. 
Given these distinctive noise characteristics and a low dimensionality of the clean image manifold, we suggest that the score functions learned from noisy data nearly point toward the clean image manifold, as illustrated in Fig.\ref{fig:teaser} 

Consequently, this implies that sampling with diffusion models trained on noisy images is likely to converge near the clean image manifold, provided the diffusion model is conditioned on a strong input image, which contains as much clean signal as possible while including as few noise as possible. This strong condition intuitively lowers the dimensionality of the manifold to be modeled and simplify the manifold geometry, encouraging the score functions learned on noisy data to point towards the clean image manifold 
The diffusion model hence learns the conditional distribution of \(p(r|x_c)\) (\(r=x'-x_c\)) and recovers the clean signal component \(x-x_c\) from \(r\).

\subsection{Conditional Diffusion Framework}
\label{sec:cond-diffusion}
As an overview, diffusion models~\cite{ho2020denoising} consist of two main steps: the forward and reverse diffusion processes. In the forward process, Gaussian noise is progressively added to the input image \( X \) over \( T \) timesteps, resulting in a fully noisy version \( X_{T} \) of \( X \). This forward process for X at timestep $t \in [0,1]$ is defined as:
\begin{equation}
    X_{t} = \sqrt{\alpha_t} X + \sqrt{1 - \alpha_t} \epsilon,
\end{equation}
where \( \epsilon \sim \mathcal{N}(0, I) \), and \( \alpha_t \) follows a predefined noise schedule that determines the amount of noise added at each timestep \( t \) in the diffusion process. The reverse diffusion process is $T$ steps of denosing with based on neural network $f_{\theta}$ where $\theta$ is network parameters. In the conditional setting, the condition $c$ is also provided to $f_{\theta}$. The overall aim is to predict the noise $\epsilon$ that is added during the forward process. The network is trained to minimize the expected distance ($L_{1}$ norm)  between the predicted noise and the actual noise across all timesteps:
\begin{equation}
    \mathcal{L}_{\text{diffusion}} = \mathbb{E}_{X, c, t} \left| f_{\theta}(X_t, c, t) - \epsilon \right|.
\end{equation}

We adapt this conditional diffusion framework to our self-supervised denoising setup. We use our noisy samples $x^{\prime}$ as inputs and their manipulated versions $x_{c}$ (see Sec. \ref{sec:condition}) as the condition. 

\subsection{Conditioning with BSN}
\label{sec:condition}

BSNs are based on the \textit{J}-invariance principle, which involves predicting target pixels by leveraging neighboring pixels while excluding the target pixels themselves. 
This implies that if the noise within the blind spot is uncorrelated with the noise in other pixels, the output will be completely noise-free. Moreover, recent advances of powerful BSN architectures have enabled the recovery of more information within the blind spot by leveraging global context \cite{wang2023lg, jang2024puca}. These properties make the outputs of BSN networks an ideal candidate for use as conditioning inputs to the diffusion model in our method. 
 

\subsection{Stabilized Reverse Diffusion Sampling}\label{sec:SRDS}
In our approach, we use the Denoising Diffusion Implicit Models (DDIM) \cite{ho2020denoising} sampling method to generate clean outputs efficiently. DDIM provides a non-Markovian sampling process, which accelerates inference by requiring fewer sampling steps than conventional diffusion models. It also removes intermediate noise injection steps during sampling, which makes the output image fully determined by the initial noise sample. We follow the exact sampling formula proposed in \cite{ho2020denoising}.
 
In addition to DDIM sampling, we introduce a novel Stabilized Reverse Diffusion Sampling (SRDS) strategy that aims to further improve the output quality in situations where the learned score function may have artifacts. These imperfections arise due to factors such as sparse, noisy sample coverage in training, as well as approximation errors in the learned score function relative to the true score function over the clean data manifold. 
 
Our SRDS approach leverages the idea of symmetry around the low-dimensional clean signal manifold by introducing symmetric noise pairs during the sampling process. The key insight here is that when symmetric noise pairs are applied to the data, the resulting denoised outputs tend to balance each other, which effectively mitigates any residual biases or inaccuracies in the learned score function. Specifically, the symmetric pairing of noise samples helps cancel out any distortions caused by approximations or sampling density limitations, producing outputs that more closely approximate the true clean data.
 
Symmetric sampling is formulated as follows. We first sample a random noise $ \epsilon \sim \mathcal{N}(0, I) $ and calculate its negation $ -\epsilon  $. For each noise sample, we execute the reverse diffusion sampling process conditioned with $x_{c}$ comes from the BSN output for T steps.
 
\begin{equation}
    \hat{x}^{\epsilon} = f_\theta(x^{\prime}, \epsilon, T), \quad \hat{x}^{-\epsilon} = f_\theta(x^{\prime}, -\epsilon, T).
\end{equation}
The final denoised output $ \hat{x} $ is obtained by averaging the predictions from two inferences:
\begin{equation}
    \hat{x} = \frac{1}{2} \left( \hat{x}^{\epsilon} + \hat{x}^{-\epsilon} \right).
\end{equation}
 
This averaging step cancels out symmetric noise artifacts introduced during inference, providing a cleaner image that better preserves critical high-frequency details (refer to Fig.\ref{fig:srds} for more details). 




\subsection{Knowledge Distillation}
\label{sec:knowledge-distillation}
Although our stabilized reverse diffusion sampling method performs well at denoising, diffusion networks are prone to hallucination due to their probabilistic nature. To address this, we incorporate a supervised denoising network and perform knowledge distillation. Specifically, we train a denoising network using noisy images paired with their denoised versions derived from SRDS predicted by our diffusion method. This approach enables us to produce deterministic results while preserving critical shape and feature information, with the additional advantage of accelerated inference. For this, we use one of the state-of-the-art supervised image restoration methods, NAF-Net \cite{nafnet}.

\section{Experiments}
\begin{figure*}
    \centering
    \includegraphics[width=0.95\linewidth]{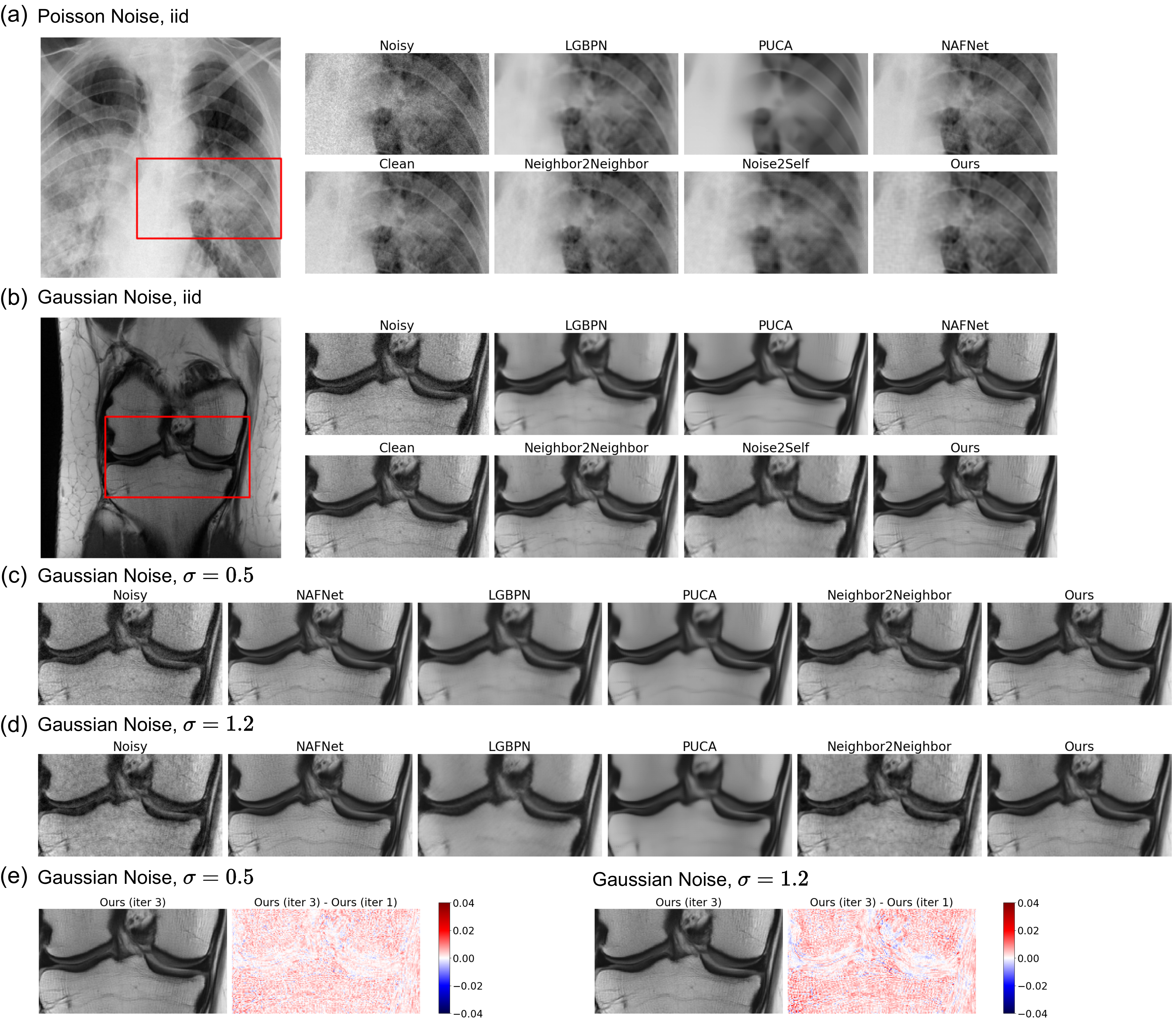}
    \caption{\textbf{Denoising results.} Visual comparisons with baseline methods on datasets with synthetic noise: (a) and (b) show pixel-wise independent denoising results for knee MRI and chest X-rays, respectively; (c) and (d) illustrate denoising for spatially correlated Gaussian noise; (e) presents denoising results after three iterations, along with a difference map highlighting changes between the first and third iterations. Overall, our method produces the sharpest outputs and visibly preserves fine structures better than the comparison methods.}
    \label{fig:iid-denoising}
\end{figure*}

\label{sec:experiments}
We designed experiments to evaluate the denoising performance and quality of our approach compared to both supervised and self-supervised SOTA image denoising models. Our evaluation aims to answer the following questions: 1) How does our model perform on pixel-wise independent synthetic noise datasets? (Sec.~\ref{sec:iid-noise}) 2) How does it perform on spatially correlated synthetic noise datasets? (Sec.~\ref{sec:corr-noise}) 3) Can it generalize to real noisy datasets? (Sec.~\ref{sec:real-world}) 4) Can multiple denoising iterations further enhance performance? (Sec.~\ref{sec:mul-iter}) 5)What is the contribution of SRDS and the knowledge distillation? (Sec.~\ref{sec:ablation})
 
\subsection{Datasets}
We evaluate our model on both synthetic noise-added medical datasets and real noisy MRI datasets.
 
\noindent \textbf{Real Datasets.} We use the T1w and FLAIR sequences of the M4Raw \cite{lyu_m4raw_2023} dataset, which contains repetitive MRI scans. For training, we use only the first scan from each sequence (assuming a single noise realization) and evaluate the model by averaging repetitive scans to generate our clean images.
 
\noindent \textbf{Synthetic Noise-Added Datasets.} Since collecting paired clean-noisy images in medical imaging is challenging, we created synthetic datasets by adding noise to existing medical imaging datasets. 1) The brain T2w and knee images from \textbf{FastMRI} dataset~\cite{fast-mri1, fastmri2}. 2) The \textbf{COVID Chest X-Ray} dataset \cite{siim-covid19-detection}, which contains chest x-ray images. For both datasets, we generate $300 \times 300$ patches and linearly normalize intensities to the range 0–1. During training, we use the default training splits and evaluate performance on the first 200 images from the validation set (for FastMRI) and test set (for COVID Chest X-Ray).
 
\noindent \textbf{Synthetic Pixel-Wise Independent Noise.} While designing these synthetic datasets, we prioritized keeping noise levels as close as possible to those encountered in contemporary medical imaging devices, with a focus on maintaining realistic PSNR dBs. For the FastMRI dataset, we added Gaussian ($\sigma=6/255$), Poisson ($\lambda=200$), Gamma ($\alpha \text{ (concentration)}=100, \beta \text{ (rate)} =100$) and for the COVID Chest X-Ray, we added Gaussian ($\sigma=6/255$), Poisson ($\lambda=700$), Gamma ($\alpha=500, \beta=500$) noises. Note that this noise addition is done only once per sample; each sample contains a single noise realization and we keep these realizations fixed for all of our experiments.



\noindent \textbf{Synthetic Spatially Correlated Noise.} To simulate spatially correlated noise, we start by generating independent and identically distributed (i.i.d.) noise using the same parameter in the last paragraph. Next, we convolve this noise with a Gaussian kernel (with $\sigma$ = 0.5 or 1.2), producing a blurred version of the noise. We then adjust the power of the blurred noise to match that of the original i.i.d. noise. Finally, we combine the adjusted blurred noise with a second, independently generated i.i.d. noise (using the same parameter) with equal weights, \(1/\sqrt{ 2}\), ensuring that the combined noise maintains the same power as the original i.i.d. noise.

\subsection{Model and Training Configuration}
For the comparison methods, we picked supervised and we train them following the original implementations, with modifications as listed below.

\noindent \textbf{Conditional Diffusion Model:} We followed the implementation of Improved Denoising Diffusion Probabilistic Models \cite{nichol2021improved} and trained our diffusion model for 200 epochs.

\noindent \textbf{Neighbor2Neighbor \cite{huang2021neighbor2neighbor}:} We used the default setup without adding additional noise on-the-fly.

\noindent \textbf{Noise2Self \cite{batson2019noise2self}:} We trained a default U-Net model for 1 epoch with a learning rate of $5 \times 10^{-4}$ using MSE loss. During inference, we additionally performed full \textit{J}-invariant reconstruction.

\noindent \textbf{LG-BPN \cite{wang2023lg}:} We adjusted the size of the blind spot to 1 for i.i.d. noise, 5 for spatially correlated noise with $\sigma=0.5$, and 9 for $\sigma=1.2$, with a learning rate of $1 \times 10^{-4}$.

\noindent \textbf{PUCA \cite{jang2024puca}:} We set the size of the blind spot to 1 for i.i.d. noise and used the default settings for other configurations, with a learning rate of $1 \times 10^{-4}$.

\noindent \textbf{BSN Selection for Conditioning}: For i.i.d. noise, we used PUCA with a blind spot size of 1. For spatially correlated noise, we used LG-BPN with the same configuration as described above.

\subsection{Synthetic Data Denoising} 
\noindent \textbf{Pixel-wise Independent Noise.} \label{sec:iid-noise} We evaluated our model on synthetic pixel-wise noise and Table~\ref{tab:result-iid} compares its performance with both supervised and self-supervised methods. Our model outperforms other baseline self-supervised methods in terms of the structural similarity index measure (SSIM). It also achieves PSNRs that are close, or higher than, Neighbor2Neighbor, the top-performing baseline method. Compared to the noisy input images, our model increases the PSNR by approximately 4 dB and is only 1 dB lower than NAF-Net, which has full access to clean images during training. Furthermore, our model achieves SSIM values close to, or sometimes higher than NAF-Net. Figure~\ref{fig:iid-denoising} illustrates the denoising results, which appear significantly sharper than those of other methods for iid noise in subfigures (a) and (b). In summary, the high PSNR dBs demonstrate the model's denoising capability, while the high SSIM indicates its ability to preserve structural similarities, which is essential for medical image denoising.

\begin{table*}[ht]
\centering
\caption{\label{tab:result-iid} Denoising performance (PSNR/SSIM) on datasets with synthetic pixel-wise iid noise. \textbf{Bold} indicates the best ($p<0.05$). }
\resizebox{\textwidth}{!}{
\begin{tabular}{c c c|c c|c c}
 & \multicolumn{2}{c|}{Brain T2w} & \multicolumn{2}{c|}{Knee} & \multicolumn{2}{c}{Chest X-ray} \\ 
 & Gauss & Poisson & Gauss & Poisson & Gauss & Poisson \\ \hline
Noisy & 32.57/0.769 & 31.34/0.849 & 32.57/0.828& 30.39/0.798 & 32.57/0.782& 31.20/0.734 \\\hline
NAFNet & 37.95/0.946 & 36.40/0.946 & 36.59/0.902 & 35.17/0.881 & 37.55/0.914 & 37.21/0.906 \\ \hline
Noise2Self & 33.00/0.895 & 32.60/0.892 & 32.26/0.751 & 31.71/0.734 & 32.93/0.784 & 33.06/0.784 \\
LG-BPN & 32.61/0.903 & 33.14/0.902 & 32.58/0.761 & 32.20/0.751 & 34.54/0.840 & 35.05/0.841 \\
PUCA & 35.45/0.909 & 34.14/0.908 & 33.82/0.797 & 31.72/0.749 & 35.56/0.856 & 34.11/0.822 \\ 
Nei2Nei & \textbf{36.69}/0.935 & \textbf{35.47}/0.930 & \textbf{36.08}/0.894 & 34.48/0.864 & \textbf{36.12}/0.894 & \textbf{35.56}/0.880 \\ \hline
Ours & \textbf{36.68}/\textbf{0.942} & 35.44/\textbf{0.942} & 36.02/\textbf{0.897} & \textbf{35.31}/\textbf{0.893} & 35.95/\textbf{0.905} & 35.30/\textbf{0.901} \\
\end{tabular}
}

\end{table*}

\noindent \textbf{Spatially Correlated Noise.} \label{sec:corr-noise}  We also hypothesize that our approach not only works with pixel-wise independent noise but also effectively handles spatially correlated noise, which is common in medical images. Table~\ref{tab:result-corr} shows that our method outperforms all self-supervised baselines in both PSNR and SSIM. Notably, we observe that other baseline methods struggle to denoise images with larger spatial correlations, generally underperforming by 1–2 PSNR dBs when comparing cases with $\sigma=0.5$ and $\sigma=1.2$. In contrast, our method maintains similar performance across both spatial correlation settings, demonstrating its robustness in handling larger correlations without significant performance degradation. Also Figure~\ref{fig:iid-denoising} (c) and (d) shows that comparison methods either over smooth the images or fail to remove noise, whereas our method successfully preserves structures while effectively denoising.

\begin{table*}[ht]
\centering
\caption{\label{tab:result-corr} Denoising performance (PSNR/SSIM) on the Knee dataset with synthetic spatially correlated noise. The numbers next to each noise type indicate the sigma values of the Gaussian kernel used to simulate spatial correlations. \textbf{Bold} indicates the best ($p<0.05$).}
\resizebox{\textwidth}{!}{
\begin{tabular}{c c c|c c|c c}
 & Gauss (0.5) & Gauss (1.2) & Poisson (0.5) & Poisson (1.2) & Gamma (0.5) & Gamma (1.2) \\ \hline
Noisy & 32.57/0.830 & 32.57/0.846 & 30.40/0.801 & 30.39/0.823 & 32.92/0.896 & 32.92/0.908 \\\hline
NAFNet & 36.41/0.900 & 36.33/0.914 & 35.33/0.898 & 35.14/0.905 & 36.84/0.913 & 36.61/0.917 \\ \hline
LG-BPN & 31.78/0.729 & 30.59/0.703 & 31.62/0.724 & 30.65/0.705 & 31.69/0.728 & 30.69/0.706\\
PUCA & 32.57/0.769 & 32.23/0.760 & 32.20/0.761 & 32.03/0.756 & 32.39/0.764 & 31.91/0.751 \\ 
Nei2Nei & 35.68/0.881 & 33.91/0.850 & 34.27/0.859 & 32.35/0.839 & 35.37/0.889 & 34.13/0.891 \\ \hline
Ours & \textbf{35.99}/\textbf{0.892} & \textbf{35.31}/\textbf{0.878} & \textbf{34.92}/\textbf{0.881} & \textbf{34.85}/\textbf{0.869} & \textbf{36.63}/\textbf{0.905} & \textbf{35.40}/\textbf{0.895}
\end{tabular}
}

\end{table*}

\subsection{Real World Denoising}
\label{sec:real-world}
We benchmark our model on a real-world brain MRI denoising dataset M4Raw \cite{lyu_m4raw_2023}. 
Table~\ref{tab:result-real} shows that our model outperforms the best-performing method, PUCA, in T1w denoising by 0.45 dB PSNR and in FLAIR denoising by 0.28 dB PSNR. Compared to the supervised model (upper-bound performance), our method underperforms only by 0.2 dB PSNR. Additionally, it outperforms NAF-Net in terms of SSIM which shows that our method is powerful in terms of preserving anatomical consistency.

\begin{table}[ht]
\centering
\caption{\label{tab:result-real}Denoising performance (PSNR/SSIM) on the M4Raw dataset. \textbf{Bold} indicates the best ($p<0.05$).}
\begin{tabular}{c c c}
& T1w & FLAIR \\ \hline
Noisy & 30.19/0.796 & 29.95/0.781 \\ \hline
NAFNet & 32.17/0.888 & 31.65/0.844 \\ \hline
Nei2Nei & 31.11/0.898 & 30.71/\textbf{0.869} \\ 
LG-BPN & 30.20/0.871 & 29.22/0.795 \\
PUCA & 31.50/0.898 & 31.16/0.852 \\ \hline
Ours & \textbf{31.95}/\textbf{0.900} & \textbf{31.44}/0.864 \\
\end{tabular}
\end{table}


\begin{table}[ht]
\centering
\caption{\label{tab:result-multi-iter}Denoising performance (PSNR/SSIM) of iterative training on the Knee (iid and spatially correlated synthetic Gaussian noise) datasets. \textbf{Bold} indicates the best ($p<0.05$).}
\begin{tabular}{c c c c }
& Knee (iid) & Knee (s=0.5) & Knee (s=1.2) \\ \hline
Noisy & 32.57/0.828 & 32.57/0.830 & 32.57/0.846 \\ \hline
iter1 & 36.02/0.897 & 35.99/0.892 & 35.31/0.878 \\
iter2 & 36.10/\textbf{0.900} & 36.05/0.898 & 35.98/0.893 \\
iter3 & \textbf{36.11}/\textbf{0.899} & \textbf{36.09}/\textbf{0.899} & \textbf{36.07}/\textbf{0.898} \\
\end{tabular}
\end{table}

\subsection{Ablation Study}
\label{sec:ablation}
\begin{figure}
    \centering
    \includegraphics[width=0.8\linewidth]{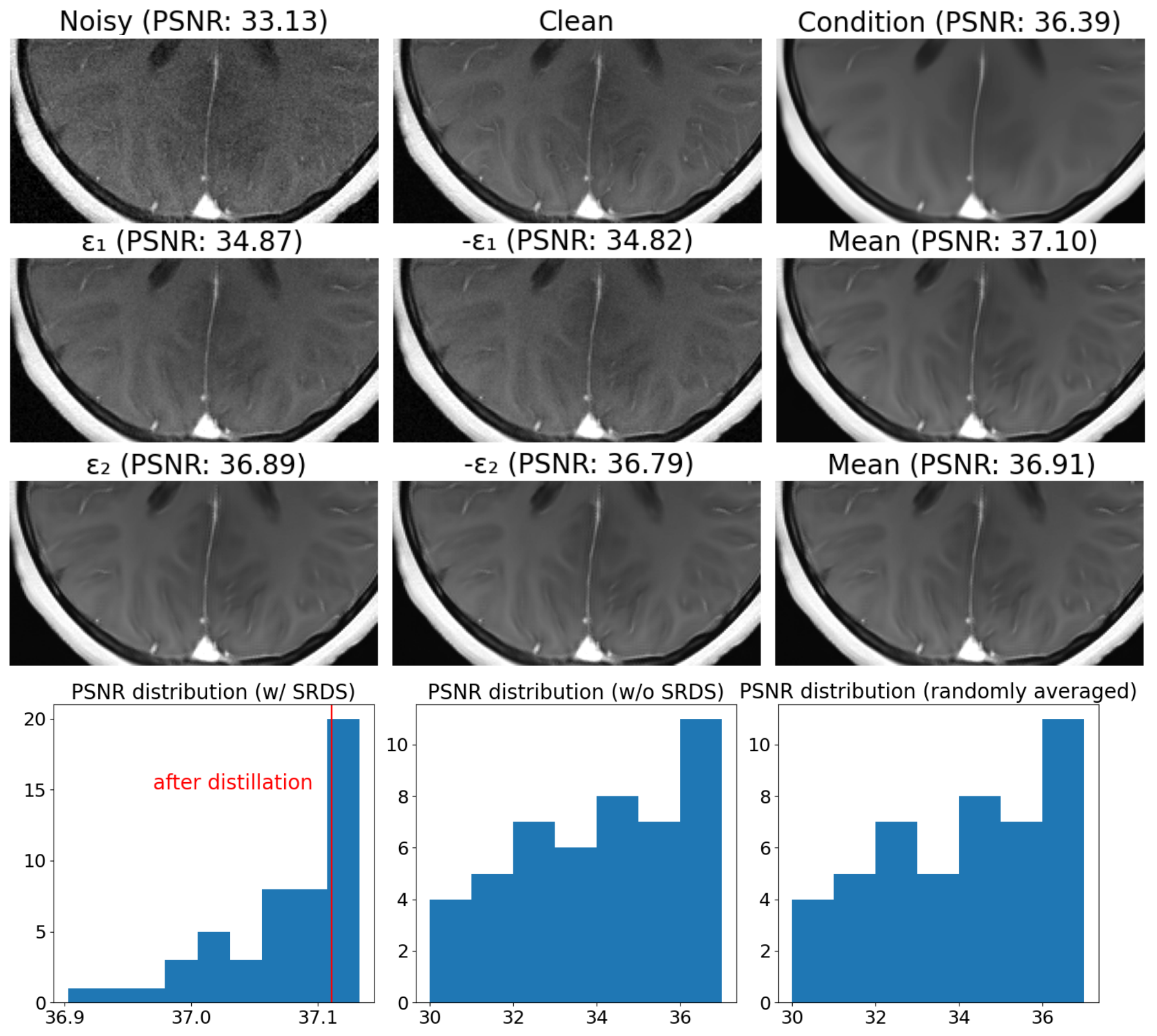}
    \caption{\label{fig:srds} \textbf{Demonstration of stabilized reverse diffusion sampling (SRDS) and knowledge distillation (KD) on one example image.} $\epsilon_{1}$ and $\epsilon_{2}$ are two random noises. Without SRDS, inference with $\epsilon_{1}$ results in a lower PSNR dB compared to inference with $\epsilon_{2}$. In contrast, SRDS enables convergence to similar PSNR dBs. The last row plots 50 random inferences with and without SRDS, as well as the average of two sampled with random noises (rather than symmetric noises in SRDS). This demonstrates that SRDS provides more stable sampling (a narrower and higher PSNR range) than both single inferences and averaging two random inferences. KD further stabilizes the inference as indicated by the red line.} 
\end{figure}

\noindent \textbf{SRDS.} The SRDS is the key component of our method. Table~\ref{tab:result-ablation} shows that using SRDS, we can improve the PSNR by 2 to 3 dB. This method is applied during the inference time and only requires another pass with the negative of the initial noise. Another observation is that without SRDS, the knowledge distillation is useless, which further decreases performance. Also, Figure~\ref{fig:srds} visualizes that SRDS  stabilizes diffusion reverse sampling procedure.

\begin{table}[!ht]
\centering

\caption{\label{tab:result-ablation}Denoising performance (PSNR/SSIM) of ablated methods SRDS and Knowledge Distillation (KD) on Knee (iid and spatially correlated synthetic Gaussian noise) datasets. \textbf{Bold} indicates the best ($p<0.05$).}
\begin{tabular}{c c | c c c }
SRDS&KD& Knee (iid) & Knee (s=0.5) & Knee (s=1.2) \\ \hline
 & & 33.48/0.793 & 33.88/0.768 & 31.50/0.729 \\
\checkmark & & 35.87/0.896 & 35.73/0.890 & 34.95/0.874 \\
 & \checkmark & 33.15/0.781 & 32.90/0.754 & 30.61/0.713 \\
\checkmark & \checkmark & \textbf{36.02}/\textbf{0.897} & \textbf{35.99}/\textbf{0.892} & \textbf{35.31}/\textbf{0.878}
\end{tabular}
\end{table}

\noindent \textbf{Knowledge Distillation.} We also investigated the effect of the final denoising network trained with knowledge distilled from our diffusion sampling procedure. Table~\ref{tab:result-ablation} shows that knowledge distillation improves the PSNR dB ranging from 0.15 to 0.36 depending on correlation size. Additionally, the last row of Figure~\ref{fig:srds} visualizes that knowledge distillation helps us to fit our predictions to stabilized high PSNR dB. Besides its performance improvement, it is important to note that the knowledge distilled network is significantly faster and deterministic, and it makes our approach applicable to real-world medical denoising problems

\section{Limitations}
\label{sec:limitations}
In our work, we focused on denoising medical images to recover fine structural details crucial for clinical diagnosis. Our experiments targeted noise levels that are commonly found in medical imaging, allowing us to tailor our approach specifically for real-world applications. However, we have not tested our model's performance on other datasets with other noise levels. Since our approach uses BSNs as the backbone model, we believe that our approach can handle datasets and noise levels that BSN networks already manage, and potentially improve BSN performance further. Another limitation of our work is the complexity of the training process. It requires three consecutive stages: first pretraining the BSN, then training the diffusion model, and finally training the knowledge-distillation network. This multi-step approach increases training time and requires a substantial amount of GPU memory due to the diffusion model component. In future studies, we will focus on optimizing training time and memory usage.

\section{Conclusion}
\label{sec:conclusion}
We proposed a self-supervised medical image denoising approach that focuses on preserving high-frequency details, an area where current SOTA models fall short. Our method introduces a novel diffusion sampling procedure that leverages symmetric noise patterns for denoising, a multi-iteration strategy to enhance performance, and knowledge distillation to make the model suitable for real-world deployment. Extensive experiments demonstrate that our model outperforms existing self-supervised SOTA approaches on both synthetic and real-world medical images. Additionally, we showed that our model generalizes well across a variety of anatomical structures and imaging modalities.




\bibliographystyle{plain}
\bibliography{main}


\end{document}